\documentclass[aps,prl,reprint,superscriptaddress]{revtex4-1}
\usepackage{helvet}
\usepackage{graphicx}
\usepackage{xspace}
\usepackage{amsmath,amssymb}
\usepackage[usenames,dvipsnames]{xcolor}
\usepackage{multibib}
\newif\ifdrafttext
\drafttexttrue 
\ifdrafttext \usepackage[colorlinks,urlcolor=black,citecolor=black,linkcolor=black]{hyperref} \else \fi

\newif\ifabstractcolor
\abstractcolorfalse

\newcounter{myfigure}
\makeatletter
\@addtoreset{figure}{myfigure}
\def\@caption@fignum@sep{~$\vert$~}%
\def\fnum@figure{\textbf{\figurename~\thefigure}}
\makeatother

\newcommand{\nuke}[1]{}
\newcommand{\mr}[1]{\ensuremath\mathrm{#1}}

\newcommand{\mb}[1]{\mathbf{#1}}
\newcommand{\uv}[1]{\mb{\hat{#1}}}

\newcommand{\note}[1]{{\color{red}{#1}}}
\newcommand{\abstracta}[1]{\textcolor{red}{#1}}
\newcommand{\abstractb}[1]{\textcolor{blue}{#1}}
\newcommand{\abstractc}[1]{\textcolor{ForestGreen}{#1}}
\newcommand{\abstractd}[1]{\textcolor{Fuchsia}{#1}}
\newcommand{\abstracte}[1]{\textcolor{Magenta}{#1}}
\newcommand{\abstractf}[1]{\textcolor{SkyBlue}{#1}}

\newcommand{\biasfield}{\ensuremath{\mathbf{B}_\mathrm{b}}}
\newcommand{\gradientfield}{\ensuremath{b_\mathrm{q}}}
\newcommand{\dbound}{\ensuremath{\uv{d}_0}}

\newcommand{\VolumeX}{\ensuremath{V}}

\newcommand*{\citen}[1]{%
  \begingroup
    \romannumeral-`\x 
    \setcitestyle{numbers}%
    \cite{#1}%
  \endgroup
}

\ifabstractcolor
\else
    \renewcommand{\note}[1]{}
    \renewcommand{\abstracta}[1]{#1}
    \renewcommand{\abstractb}[1]{#1}
    \renewcommand{\abstractc}[1]{#1}
    \renewcommand{\abstractd}[1]{#1}
    \renewcommand{\abstracte}[1]{#1}
    \renewcommand{\abstractf}[1]{#1}
\fi

\renewcommand{\figurename}{Figure}

\begin{document}

\title{Tying Quantum Knots*}

\author{D.~S.~Hall}
\affiliation{Department of Physics and Astronomy, Amherst College, Amherst, Massachusetts 01002--5000, USA}

\author{M. W. Ray}
\affiliation{Department of Physics and Astronomy, Amherst College, Amherst, Massachusetts 01002--5000, USA}

\author{K.~Tiurev}
\affiliation{QCD Labs, COMP Centre of Excellence, Department of Applied Physics, Aalto University, P.O. Box 13500, FI--00076 Aalto, Finland}

\author{E.~Ruokokoski}
\affiliation{QCD Labs, COMP Centre of Excellence, Department of Applied Physics, Aalto University, P.O. Box 13500, FI--00076 Aalto, Finland}

\author{A. H.~Gheorghe}
\affiliation{Department of Physics and Astronomy, Amherst College, Amherst, Massachusetts 01002--5000, USA}

\author{M.~M\"ott\"onen}
\affiliation{QCD Labs, COMP Centre of Excellence, Department of Applied Physics, Aalto University, P.O. Box 13500, FI--00076 Aalto, Finland}


\date{\today}

\begin{abstract}
\abstracta{Knots are familiar entities that appear at a captivating nexus of art, technology, mathematics, and science~\cite{Adams1994}. As topologically stable objects within field theories, they have been speculatively proposed as explanations for diverse persistent phenomena, from atoms and molecules~\cite{Thomson1867} to ball lightning~\cite{Ranada1996} and cosmic textures in the universe~\cite{Cruz2007}.} \abstractb{Recent experiments have observed knots in a variety of classical contexts, including nematic liquid crystals~\cite{Smalyukh2009,Tkalec2011,Sec2014}, DNA~\cite{Han2010}, optical beams~\cite{Leach2005,Dennis2010}, and water~\cite{Kleckner2013}.} \abstractc{However, no experimental observations of knots have yet been reported in quantum matter.} \abstractd{We demonstrate here the controlled creation~\cite{Kawaguchi2008} and detection of knot solitons~\cite{Faddeev1997,Battye1998} in the order parameter of a spinor Bose--Einstein condensate.} \abstracte{The experimentally obtained images of the superfluid directly reveal the circular shape of the soliton core and its accompanying linked rings. Importantly, the observed texture corresponds to a topologically non-trivial element of the third homotopy group~\cite{Nakahara2003} and demonstrates the celebrated Hopf fibration~\cite{Hopf1931}, which unites many seemingly unrelated physical contexts~\cite{Urbantke2003,Moore2008}.} \abstractf{Our observations of the knot soliton establish an experimental foundation for future studies of their stability and dynamics within quantum systems~\cite{Hietarinta2004}.} 

\end{abstract}


\maketitle

Knots are defined mathematically as closed curves in three-dimensional space~\cite{Adams1994}. A trivial example is a circle, which is also known as an unknot. More complicated knots have been extensively classified and tabulated by determining whether they can be continuously deformed, one into another, without permitting the curve to pass through itself. Although nontrivial knots are commonly associated with physical strings, they can also appear in line-like vortices, the cores of which trace the closed curves. A celebrated example is Kelvin's early atomic theory, which is linked to the existence and dynamics of knotted vortex rings in an ethereal fluid~\cite{Thomson1867}. More recently, nontrivial vortex knots have been created and identified experimentally in water~\cite{Kleckner2013} and optical beams~\cite{Leach2005,Dennis2010}, and discussed theoretically in the context of superfluid turbulence~\cite{Barenghi2007}.

\begin{figure}[htb!]
\ifdrafttext\includegraphics[width=\linewidth]{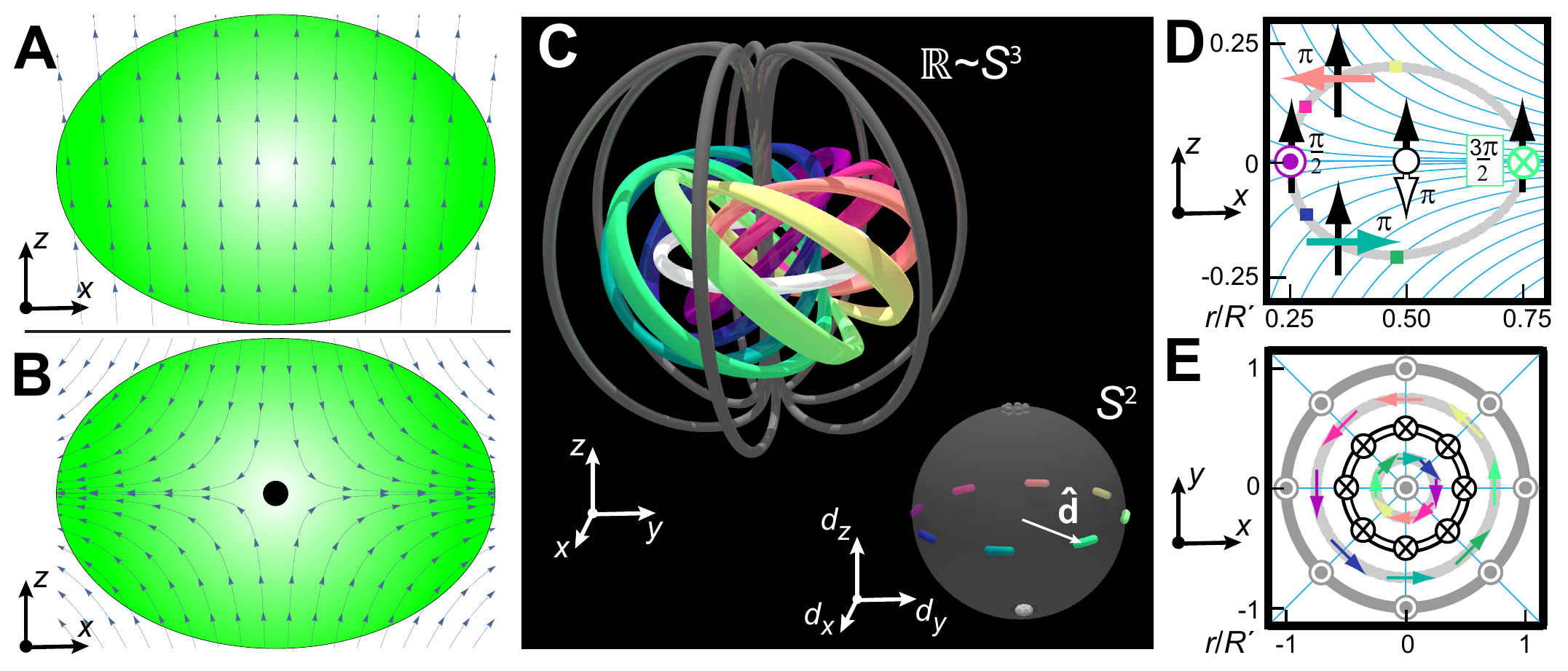} \else \fi %
\caption{\label{fig:introduction}\textbf{Structure of the knot soliton and the method of its creation.} \textbf{a},\textbf{b}, Schematic magnetic field lines before (a) and during (b) the knot formation, with respect to the condensate (green ellipse). \textbf{c}, Knot soliton configuration in real space and its relation to the nematic vector $\uv{d}$ in $S^2$ (inset). The inner white ring ($d_z=-1$, $m=0$) is the core of the knot soliton. The surrounding coloured bands ($d_z=0$, $m=\pm 1$) define the surface of a torus, with colours representing
the azimuthal angle of $\uv{d}$ which winds by $2\pi$ in both the toroidal and poloidal directions. The outer dark grey rings ($d_z=1$, $m=0$) indicate the boundary of the soliton. \textbf{d,e}, When tying the knot, the initially $z$-pointing nematic vector (black arrows) precesses about the direction of the local magnetic field (cyan lines) to achieve the final configuration (coloured arrows). The dashed grey line shows where $d_z=0$, the white line indicates the soliton core ($d_z=-1$), and the dark grey line defines the boundary of the volume ($d_z=1$).}
\end{figure} 


\begin{figure}[htb!]
\ifdrafttext\includegraphics[width=\linewidth]{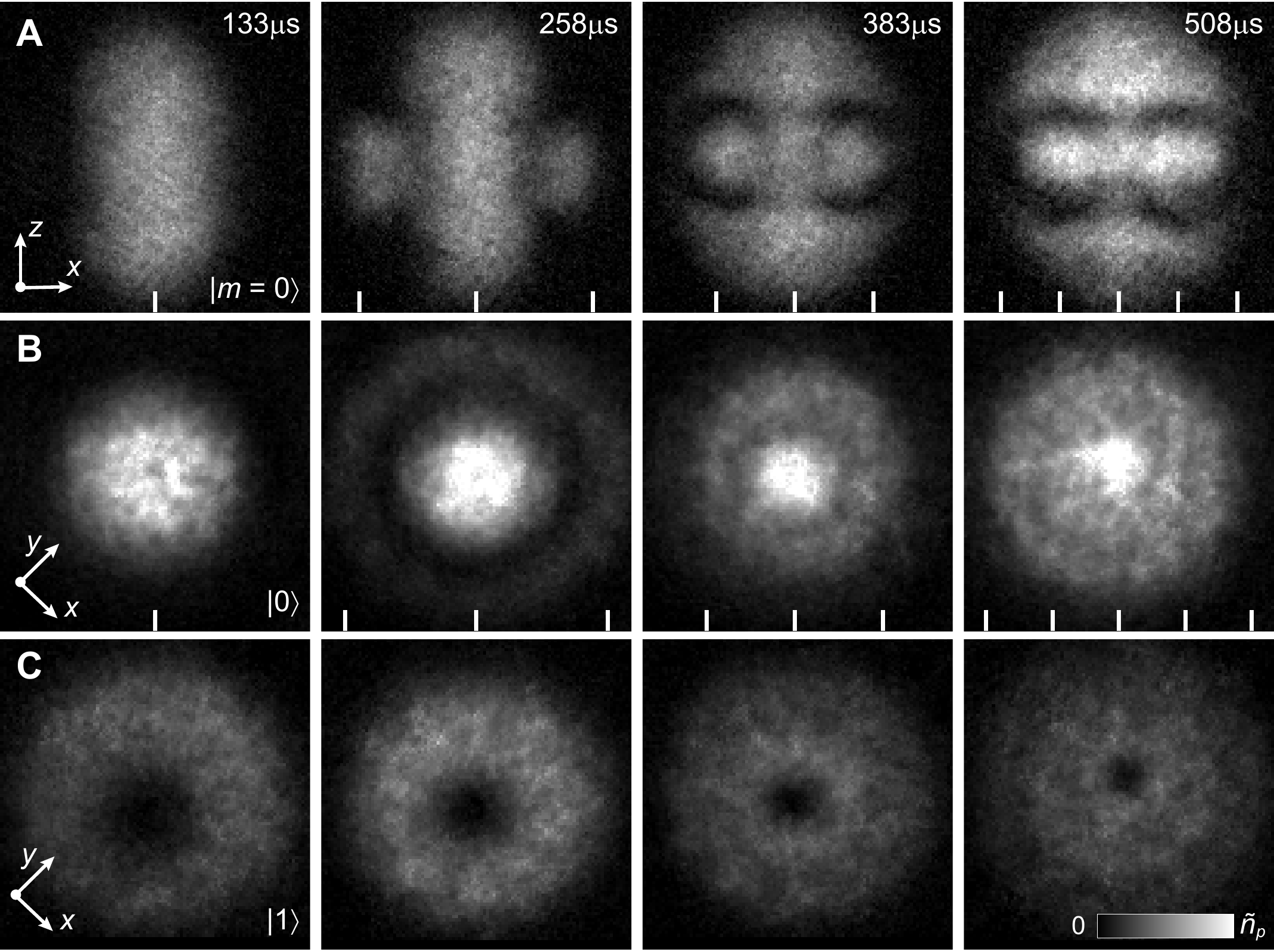} \else \fi %
\caption{\label{fig:evolution}\textbf{Tying the knot soliton by winding the nematic vector.} Experimental side (\textbf{a}) and top (\textbf{b,c}) images of the atomic column density of the $m=0$ (\textbf{a,b}) and $m=-1$ (\textbf{c}) spinor components at the indicated evolution times. Continuous rotation of the nematic vector brings the knot soliton into the condensate through its boundary, where the directors rotate by $2\pi$ in $\sim 440~\mu$s. The intensity peaks inside the circular intensity dips (a) show the core of the knot soliton. The dips correspond to the colourful torus shown in Fig.~\ref{fig:introduction}c occupied by the $m=\pm 1$ components (see also Fig.~\ref{fig:best_knot}). The analytical locations of the core and regions for which $d_z=1$ [see equation~\eqref{eq:volumeR}] are shown as ticks on the horizontal axes.
 For (\textbf{a}) the field of view is $246~\mu\mathrm{m} \times 246~\mu\mathrm{m}$ and the maximum pixel intensity corresponds to column densities in excess of $\tilde{n}_p = 8.5 \times 10^{8}$~cm$^{-2}$; for (\textbf{b},\textbf{c}) these quantities are respectively $219~\mu\mathrm{m} \times 219~\mu\mathrm{m}$ and $\tilde{n}_p = 1.0 \times 10^{9}$~cm$^{-2}$.}
\end{figure} 


\begin{figure}[htb!]
\ifdrafttext\includegraphics[width=\linewidth]{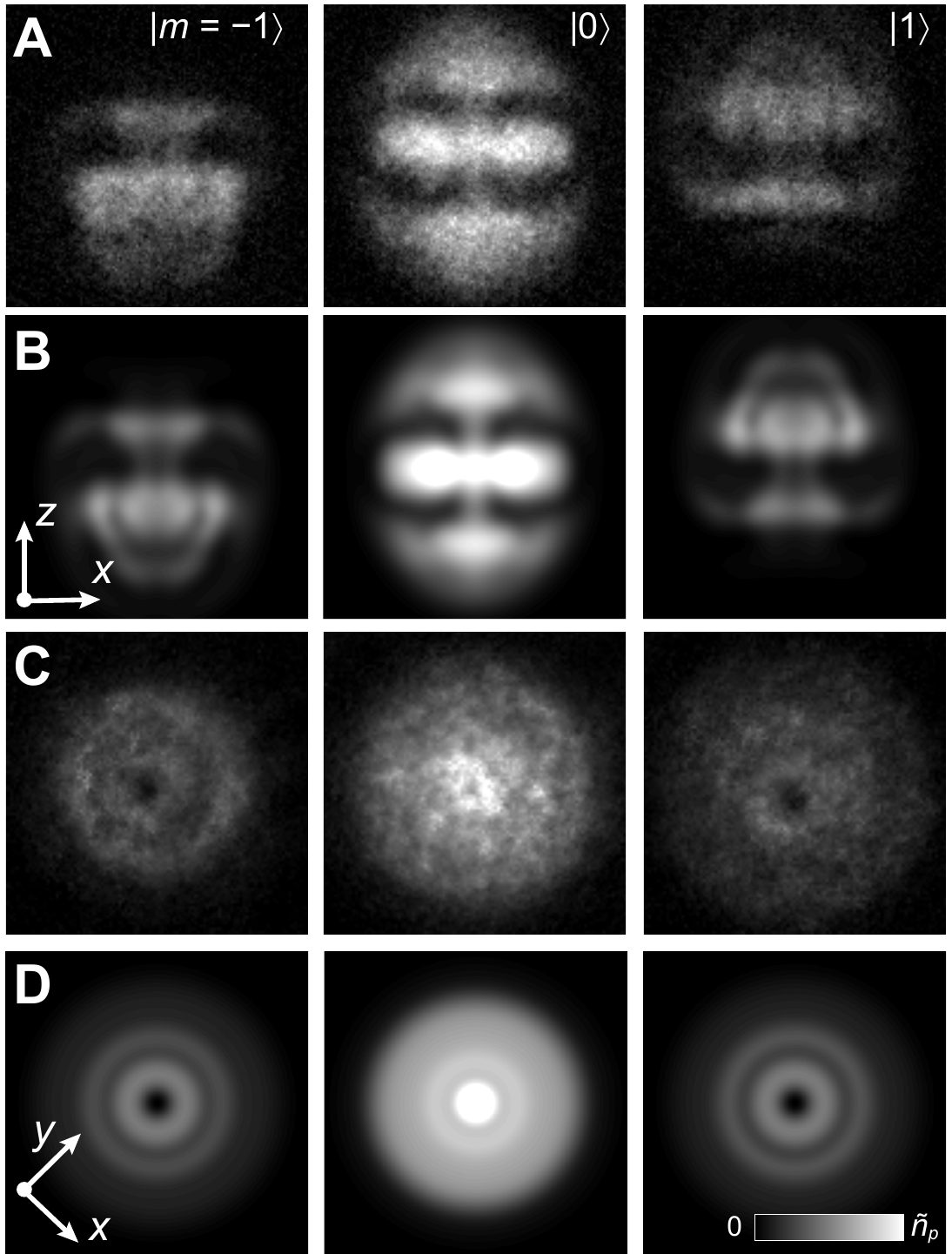} \else \fi %
\caption{\label{fig:best_knot}\textbf{Comparison of experiment with theory.} Side (\textbf{a},\textbf{b}) and top (\textbf{c},\textbf{d}) images of the experimentally (\textbf{a},\textbf{c}) and theoretically (\textbf{b},\textbf{d}) obtained atomic column densities in all different spinor components as indicated. The number of particles is $2.4\times 10^5$, and the knot is tied for $T_\mathrm{evolve}=558$~$\mu$s. For (\textbf{a},\textbf{b}) the field of view is $246~\mu\mathrm{m} \times 246~\mu\mathrm{m}$ and the maximum pixel intensity corresponds to column densities in excess of $\tilde{n}_p = 8.5 \times 10^{8}$~cm$^{-2}$; for (\textbf{c},\textbf{d}) these quantities are $219~\mu\mathrm{m} \times 219~\mu\mathrm{m}$ and $\tilde{n}_p = 1.0 \times 10^{9}$~cm$^{-2}$, respectively.}
\end{figure} 


\begin{figure}[htb!]
\ifdrafttext\includegraphics[width=\linewidth]{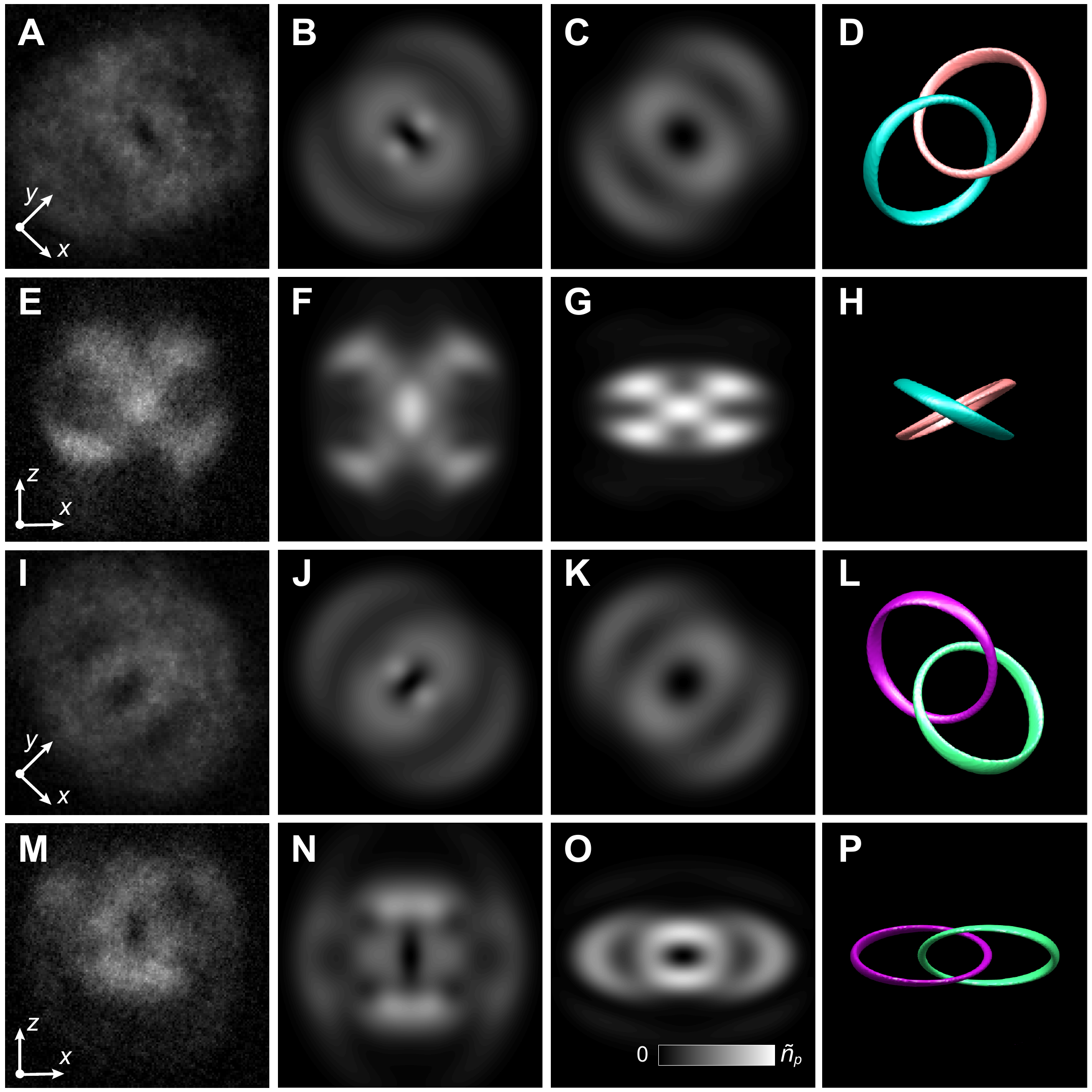} \else \fi %
\caption{\label{fig:links}\textbf{Linked preimages.} Experimental (\textbf{a}) and simulated (\textbf{b}) top images of the $m=0$ spinor component for $T_\mathrm{evolve}=508~\mu$s and projections along $-x$, where the maximum pixel intensity corresponds to column densities in excess of $\tilde{n}_p = 1.0\times 10^{9}~\mathrm{cm}^{-2}$ and the field of view $219~\mu\mathrm{m} \times 219~\mu\mathrm{m}$. \textbf{c}, Simulated top image of the condensate prior to expansion, with $\tilde{n}_p = 2.6\times 10^{11}~\mathrm{cm}^{-2}$. Projection along $\alpha\in\{\pm x,\pm y\}$ results in a column density with pronounced intensity along the preimages of $d_\alpha=1$ and $d_\alpha=-1$. \textbf{d}, Preimages of $d_x=\pm 1$ from the simulation of panel (c), with colours corresponding to those of Fig.~\ref{fig:introduction}c. The field of view in (c,d) is $15~\mu\mathrm{m} \times 15~\mu\mathrm{m}$. \textbf{e--h}, Same as (a--d), but for images taken from the side. The field of view in (e,f) is $246~\mu\mathrm{m} \times 246~\mu\mathrm{m}$ with $\tilde{n}_p = 8.5\times 10^{8}~\mathrm{cm}^{-2}$, and the field of view in (g,h) is the same as in (c,d). \textbf{i--l}, Same as (a--d), but for projection along $y$ and preimages $d_y=\pm 1$. \textbf{m--p}, Same as (e--h), but for projection along $y$ and preimages $d_y=\pm 1$.}
\end{figure} 

Knots can also appear as particle-like solitons in classical and quantum fields~\cite{Manton2004}, the nature of which has been a subject of intense mathematical interest for more than eighty years~\cite{Hopf1931}. In this case, the closed curve is a ring that describes the core of the soliton. This ring is surrounded by an infinite number of similar rings, each linked with all of the others to generate a knotted field structure~\cite{Radu2008}. For example, the classical Maxwell's equations admit solutions that involve knot solitons in which the rings are electric and magnetic field lines~\cite{Ranada1990}. Quantum-mechanical examples have been theoretically proposed~\cite{Faddeev1997,Battye1998} for the Faddeev--Skyrme model, in which each of the linked rings consists of the points in space sharing a particular direction of the field.

In general, knot solitons are non-singular topological excitations~\cite{Nakahara2003} that change smoothly and non-trivially in all three spatial dimensions.
They are therefore described by the third homotopy group, $\pi_3$, which classifies such textures according to whether or not they can be continuously transformed into one another. One-dimensional solitons and singular vortex lines, both belonging to the fundamental group $\pi_1$, have been identified experimentally in superfluids~\cite{Denschlag2000,Burger1999,Vinen1961}, as have two-dimensional skyrmions~\cite{Choi2012} and singular monopoles~\cite{Ray2015} belonging to the second homotopy group, $\pi_2$. Since singular defects within $\pi_3$ are unrealisable monopoles in four spatial dimensions, the knot soliton is an instance of the only general texture type that has not yet been identified experimentally within a medium described by a quantum-mechanical order parameter.

In this Letter we demonstrate the creation and observation of knot solitons in the polar phase of a spinor Bose--Einstein condensate. We adopt the theoretical method proposed in ref.~\citen{Kawaguchi2008} and implement it using experimental techniques that have recently been used to create Dirac monopoles~\cite{Ray2014} and isolated monopoles~\cite{Ray2015}. An overview of the experiment is given in Fig.~\ref{fig:introduction}. The core of the knot soliton is first observed as a ring of enhanced particle density at the periphery of the condensate that shrinks inwards as time advances (Fig.~\ref{fig:evolution}). The presence of the soliton is confirmed by its excellent agreement with analytical theory and with numerical simulations (Fig.~\ref{fig:best_knot}). Strikingly, we directly image the linked structure of the knot soliton, as shown in Fig.~\ref{fig:links}.

The order parameter describing a spin-1 Bose--Einstein condensate may be written as
\begin{align}
\Psi(\mb{r},t) = \sqrt{n(\mb{r},t)}e^{i\phi(\mb{r},t)}\zeta(\mb{r},t)
\end{align}
where $n$ is the atomic density, $\phi$ is a scalar phase, and $\zeta=(\zeta_{+1},\,\zeta_0,\,\zeta_{-1})^{\mr{T}}$ is a three-component $z$-quantized spinor with $\zeta_m=\langle m | \zeta \rangle$. We restrict our attention here to the polar phase of the condensate, which is obtained by spin rotations
\begin{align}
\mathcal{D}(\alpha,\beta)\begin{pmatrix}0 \\ 1 \\ 0\end{pmatrix} = \frac{1}{\sqrt{2}}\begin{pmatrix}
-e^{-i\alpha}\sin\beta \\ \sqrt{2}\cos\beta \\ e^{i\alpha}\sin\beta
\end{pmatrix} = \frac{1}{\sqrt{2}} \begin{pmatrix}
-d_{x} + id_{y} \\ \sqrt{2}d_{z} \\ d_{x} + id_{y} \end{pmatrix} \label{eq:polard}
\end{align}
for angles $\beta$ and $\alpha$ about the $y$ and $z$ axes, respectively.
The polar order parameter may therefore be expressed as
\begin{align} \label{eq:orderparameter}
\Psi(\mb{r},t) = \sqrt{n(\mb{r},t)}e^{i\phi(\mb{r},t)}\uv{d}(\mb{r})
\end{align}
in terms of the nematic vector, $\uv{d}$, defined by equation~\eqref{eq:polard}.

The nematic vector field $\uv{d}(\mb{r})$ maps points 
in real space $\mb{r} \in \mathbb{R}^3$ to points on the surface of the unit sphere $\uv{d} \in S^2$. In our case, the nematic vector assumes a constant value, $\dbound$, at the boundary of a certain volume $\VolumeX$. We restrict our studies to textures inside the volume $\VolumeX$, which as a result can be identified with $S^3$, the surface of a four-dimensional ball. Nontrivial mappings $\uv{d}(\mb{r})$ from $S^3$ to $S^2$ lead to knotted field configurations characterised by integer topological charges or Hopf invariants~\cite{Manton2004}, $Q$, as determined by the third homotopy group $\pi_3(S^2)=\mathbb{Z}$. Field configurations with different Hopf invariants cannot be continuously deformed into one another and are therefore topologically distinct.

Taken together, the points in $\VolumeX$ at which $\uv{d}$ assumes the same value, $\uv{d}_\mathrm{c}$, define a closed curve known as the preimage of $\uv{d}_\mathrm{c}$. Each of these preimages is linked with all of the others, which are associated with different $\uv{d}$, exactly $Q$ times. Thus the linking number is equivalent to the Hopf invariant~\cite{Manton2004}, and provides an alternative perspective on its physical significance.

\begin{figure}[p!]
\ifdrafttext\includegraphics[width=\linewidth]{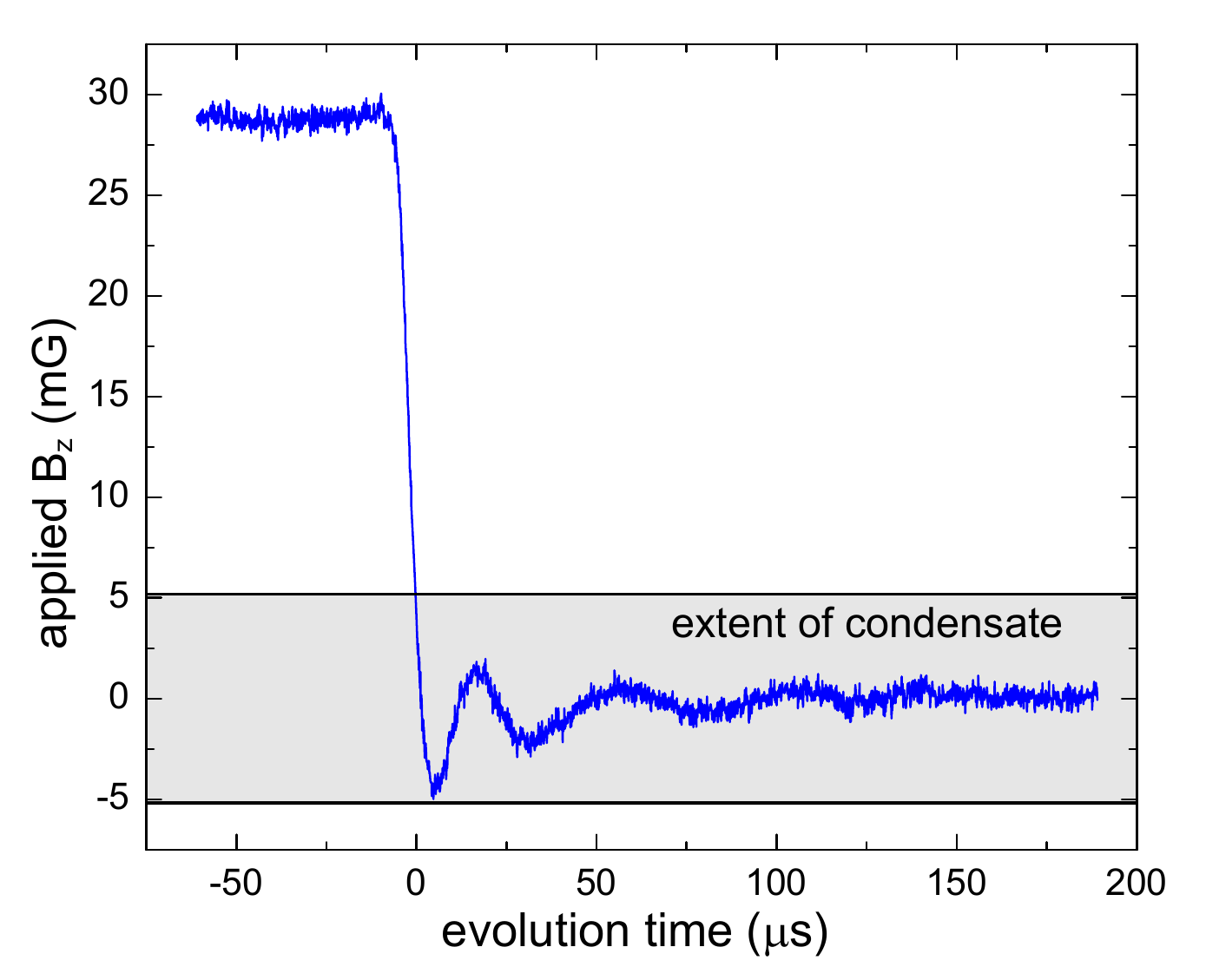} \else \fi %
\caption{\label{fig:fastramp}\textbf{The rapid change to the applied magnetic field.} {Typical averaged trace of the change in the applied current to the magnetic field coils that initiates the knot creation process. The vertical axis is expressed in terms of $B_z$ using coil calibration data obtained from microwave spectroscopy. Some field contributions from external sources, such as eddy currents, are not included. The grey region shows the effective extent of the condensate as determined by the value of the magnetic field gradient.}}
\end{figure}

In this experiment we consider the Hopf map~\cite{Hopf1931}, which has $Q=1$ and is generated physically in our system by spin rotations in an inhomogeneous magnetic field. We begin with an optically trapped $^{87}$Rb condensate (see Methods) described by the spinor $\xi=(0,1,0)^T$, corresponding to $\uv{d}=\dbound=\uv{z}$. The inhomogeneous magnetic field is given by
\begin{align}
\mb{B}(\mathbf{r'},t)=\gradientfield(x'\uv{x}+y'\uv{y}-z'\uv{z})+\biasfield(t)
\end{align}
where the condensate is taken to be at the origin of the rescaled coordinate system $x'=x$, $y'=y$, and $z'=2z$. In the gradient $\gradientfield =4.5$~G/cm and effective bias field $B_z \sim 30$~mG, the zero point of the magnetic field is initially 33~$\mu$m away from the centre of the condensate. The creation of the knot begins with a sudden change of $\biasfield(t)$ that places the field zero at the centre of the condensate, ideally leaving its state unchanged (see Fig.~\ref{fig:introduction}a,b and Extended Data Fig.~\ref{fig:fastramp}). The nematic vectors then precess about the direction of the local magnetic field at their spatially-dependent Larmor frequencies
\begin{align}\label{eq:Larmor}
\omega_\textrm{L}(\mb{r'}) = \frac{g_\textrm{F} \mu_\textrm{B} |\mb{B}(\mathbf{r'},t)|}{\hbar} = \frac{g_\textrm{F} \mu_\textrm{B} \gradientfield r'}{\hbar}
\end{align}
where $g_\textrm{F}$ is the atomic Land\'e $g$-factor, $\mu_\textrm{B}$ is the Bohr magneton, and $r'=\sqrt{x'^2+y'^2+z'^2}$. The optimal result is the time-dependent nematic vector field
\begin{align}
\uv{d}(\mb{r'})=\exp\left[-i \omega_\textrm{L}(\mb{r'})t\,\uv{B}(\mb{r'})\cdot \mb{F} \right]\dbound
\label{eq:nematic}
\end{align}
where $\mb{F}$ is the vector of dimensionless spin-1 matrices in the Cartesian basis. Importantly, $\uv{d}(\mb{r'})=\dbound$ for all points satisfying $\omega_L(\mb{r'})t = 2\pi$, and hence we may choose the volume $\VolumeX$ to be a ball of radius
\begin{align}\label{eq:volumeR}
R'=\frac{2\pi\hbar}{g_F\mu_\textrm{B}\gradientfield t}.
\end{align}

Figure~\ref{fig:introduction}d--e illustrates how the nematic vector assumes its knot soliton configuration as a result of the spatially dependent Larmor precession. The core of the soliton is identified with the preimage of the south pole of $S^2$, i.e., $\uv{d}_\mathrm{core}=-\dbound$,
which lies in the $x'y'$-plane. This ring defines a circle (Fig.~\ref{fig:introduction}c--e) in which the condensate is entirely in the $m=0$ spinor component [equation~\eqref{eq:polard}]. The comparable preimage of the north pole, $\uv{d}=\dbound$, includes the $z'$-axis and all points on the boundary of $\VolumeX$. Because antipodal points on $S^2$ correspond to the same spinor up to a sign [equation~\eqref{eq:orderparameter}], this preimage is also entirely in the $m=0$ component. The preimages of the equatorial points on the two-sphere consist of linked rings that define a toroidal tube enclosing the core, as shown in Fig.~\ref{fig:introduction}c. Since $d_z=0$ at the equator of $S^2$, this torus consists of overlapping $m=\pm 1$ components [equation~\eqref{eq:polard}]. Elsewhere, $\uv{d}$ smoothly interpolates between these values.

After an evolution time $T_\mathrm{evolve}$ we apply a projection ramp in which the bias field $B_z$ is rapidly changed to move the field zero far from the centre of the condensate~\cite{Ray2014,Ray2015}. The condensate is then released from the optical trap, whereupon it expands and falls under the influence of gravity. Subsequently, its spinor components are separated and imaged simultaneously along both the vertical ($z$) and horizontal ($y$) axes.

The temporal evolution of the particle column densities in the $m=0$ component, $\int nd^2_z\,dy$, is shown in Fig.~\ref{fig:evolution}a,b. The pictures show the combined preimages of the poles of $S^2$ ($d_z=\pm 1$), revealing in one picture both the core of the knot and the boundary of $\VolumeX$. The preceding analytical result for the radius of the core, $R'/2$ from equation~\eqref{eq:volumeR}, agrees well with the experimental observations.

\begin{figure}[p!]
\ifdrafttext\includegraphics[width=\linewidth]{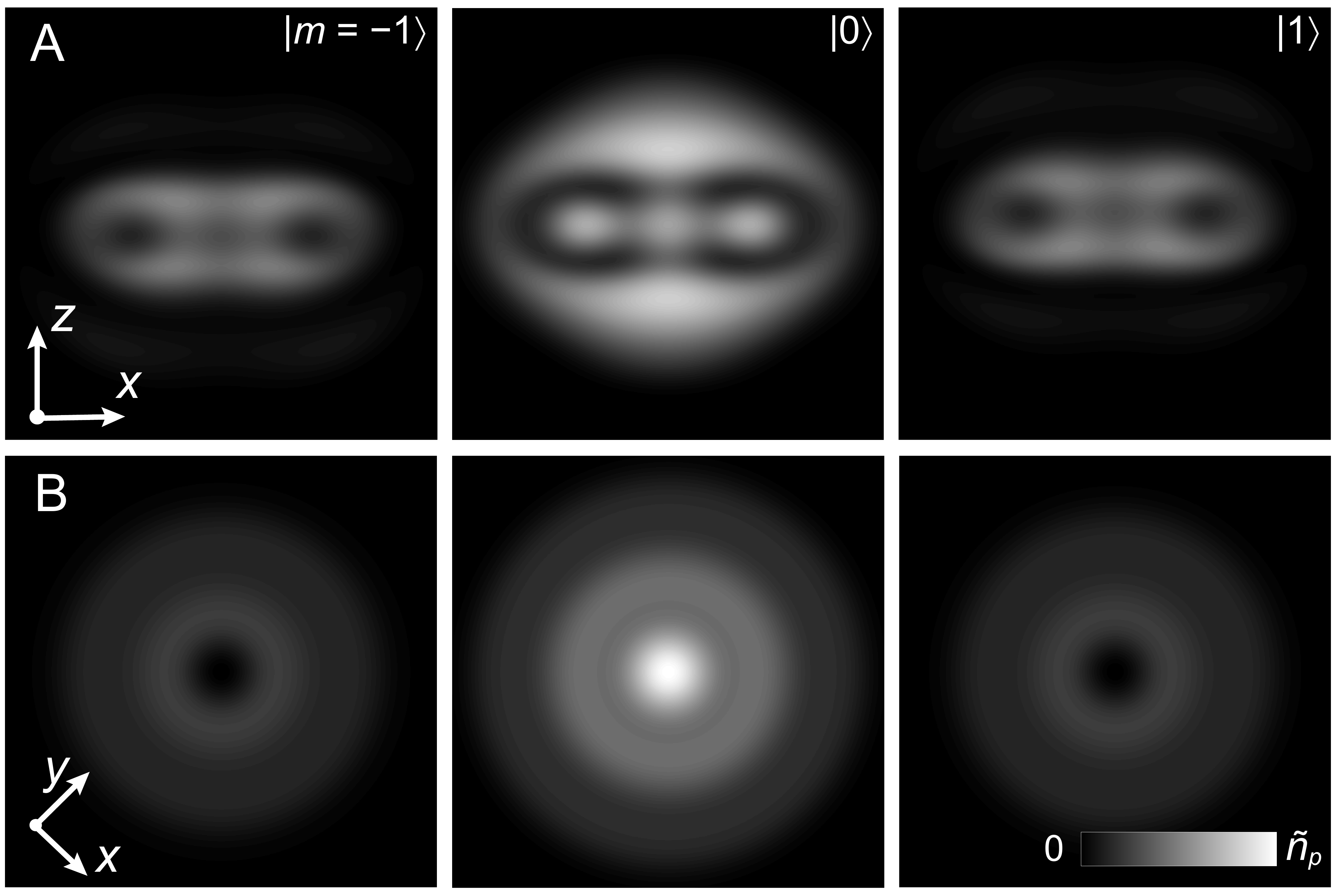} \else \fi %
\caption{\label{fig:preexpansion}\textbf{Numerical simulation of the knot creation before expansion.} Horizontally (\textbf{a}) and vertically (\textbf{b}) integrated particle densities of a condensate just before the projection ramp after an evolution time of 557~$\mu$s, with matching parameters as in Fig.~\ref{fig:best_knot}. The field of view is $15~\mu\mathrm{m} \times 15~\mu\mathrm{m}$ in each frame, and the maximum pixel intensity corresponds to $\tilde{n}_p = 3.8 \times 10^{11}$ cm$^{-2}$.} 
\end{figure}

Figure~\ref{fig:best_knot} provides a detailed comparison of the experimentally obtained knot soliton with numerical simulations of the corresponding Gross--Pitaevskii equation (see Methods) with no free parameters. The very good correspondence between the experiment and the simulation, together with the qualitatively correct behaviour of the $m=\pm 1$ spinor components that jointly accumulate in the vicinity of the intensity minima of the $m=0$ component, provide further evidence that the observed texture is that of a knot soliton. Note that the $m=\pm 1$ components do not overlap as a result of the time-of-flight imaging technique (see Extended Data Fig.~\ref{fig:preexpansion} for simulated images of the spinor components prior to expansion).

By definition the nematic vector is aligned with the local spin quantization axis, along which the condensate is fully in the $m=0$ component [see equation~\eqref{eq:polard}]. Remarkably, a projection ramp taken along an arbitrary axis, $\eta$, populates the $m=0$ component with the preimages of the antipodal points in $S^2$ corresponding to $d_\eta = \pm 1$. By performing the projection ramp along $x$ and $y$, for example, we can observe the linked preimages of $d_x=\pm 1$ and $d_y=\pm 1$, respectively, in the $m=0$ component  (Fig.~\ref{fig:links}). These images explicitly demonstrate the linked rings of the Hopf fibration and provide conclusive evidence of the existence of the knot soliton.

Our observations suggest future experiments on the dynamics, stability, and interactions of knot solitons~\cite{Hietarinta2004}. Experimental creation of multiply-charged and knotted-core solitons in quantum fields stands as another promising research direction. Furthermore, stabilising the knot soliton against dissipation, a feature associated with textures in the Faddeev--Skyrme model~\cite{Faddeev1997,Battye1998}, remains an important experimental milestone.


\begin{center}
\textbf{METHODS}
\end{center}

{\bf Experiment.} The condensate initialisation, trapping, and imaging techniques are essentially identical to those of ref.~\onlinecite{Ray2015}. The key technical difference in the present experiments is that we bring the magnetic field zero suddenly into the condensate centre, in contrast to the adiabatic creation ramp in ref.~\onlinecite{Ray2015}. Extended Data Fig.~\ref{fig:fastramp} shows the measured temporal evolution of the electric current controlling $B_z$ during its excursion, expressed in units of the magnetic field.  We define $t=0$ to be the moment at which the field zero has traversed 90\% of the distance towards its final location at the centre of the condensate. The strength of the quadrupole gradient field is estimated by repeating the knot creation experiment with a small bias field offset applied along the $x$-axis, which introduces a fringe pattern that winds at a rate proportional to the strength of the gradient.

The crossed-beam optical dipole trap operates at 1064~nm with frequencies $\omega_r \sim 2\pi \times 130$~Hz and $\omega_z \sim 2\pi \times 170$~Hz in the radial and axial directions, respectively. The total number of particles in the condensate at the moment of imaging is typically $2.5\times 10^5$.

{\bf Data.} The experimentally obtained images of knot solitons shown in this manuscript represent typical results selected from among several hundred successful realisations taken under similar conditions over the course of more than a year. Remarkably, almost identical knot solitons have been created with several minutes of time elapsed between the realisations while not changing the applied control sequences.

{\bf Simulation.} We theoretically describe the low-temperature dynamics of the condensate using the full three-dimensional spin-1 Gross--Pitaevskii equation
\begin{align}
i\hbar\partial_t\Psi(\mathbf{r})=\{h(\mathbf{r})+n(\mathbf{r})[c_0+c_2\mathbf{S}(\mathbf{r})\cdot\mathbf{F}]-i\Gamma n^2(\mathbf{r})\}\Psi(\mathbf{r})
\end{align}
where we denote the single-particle Hamiltonian by $h(\mathbf{r})$, the spin vector by $\mathbf{S}(\mathbf{r})=\zeta(\mathbf{r})^\dagger \mathbf{F}\zeta(\mathbf{r})$, and the density--density and spin--spin coupling constants by $c_0=4\hbar^2(a_0+2a_2)/(3m)$ and $c_2=4\hbar^2(a_2-a_0)/(3m)$, respectively. We employ the literature values for the three-body recombination rate $\Gamma=2.9\times\hbar\times 10^{-30}$~cm$^6$/s, the $^{87}$Rb mass $m=1.443\times 10^{-25}$~kg, and the $s$-wave scattering lengths $a_0=5.387$~nm and $a_2=5.313$~nm. The single-particle Hamiltonian assumes the form $h(\mathbf{r})=-\hbar^2\nabla^2/(2m)+V_\textrm{opt}(\mathbf{r})+g_\textrm{F}\mu_\textrm{B}\mathbf{B}(\mathbf{r},t)\cdot\mathbf{F}+q[\mathbf{B}(\mathbf{r},t)\cdot\mathbf{F}]^2,$ where the strength of the quadratic Zeeman effect is given by $q=2\pi\hbar\times 2.78$~MHz/T and the optical trapping potential is approximated by $V_\textrm{opt}(\mathbf{r})=m\omega_r^2(x^2+y^2)+m\omega_z^2z^2.$ The Gross--Pitaevskii equation is integrated using a split-operator method and fast Fourier transforms on a discrete grid of size $8 \times 10^6$. The computations are carried out using state-of-the-art graphics processing units. The simulations reproduce the experimental results with no free parameters: Only literature values for constants and independently measured parameters, such as the temporal dependence of the magnetic field, are employed. The magnetic field gradient that is briefly applied to separate the different spinor components during the time-of-flight imaging is not included in the simulations.


\bibliographystyle{naturemag}

\begin{thebibliography}{10}
\expandafter\ifx\csname url\endcsname\relax
  \def\url#1{\texttt{#1}}\fi
\expandafter\ifx\csname urlprefix\endcsname\relax\def\urlprefix{URL }\fi
\providecommand{\bibinfo}[2]{#2}
\providecommand{\eprint}[2][]{\url{#2}}

\bibitem{Adams1994}
\bibinfo{author}{Adams, C.~C.}
\newblock \emph{\bibinfo{title}{The Knot Book}} (\bibinfo{publisher}{W. H.
  Freeman and Co.}, \bibinfo{address}{New York, USA}, \bibinfo{year}{1994}).

\bibitem{Thomson1867}
\bibinfo{author}{Thomson, W.}
\newblock \bibinfo{title}{On vortex atoms}.
\newblock \emph{\bibinfo{journal}{Proc. Roy. Soc. Edinburgh}}
  \href{http://dx.doi.org/10.1017/S0370164600045430}{\textbf{\bibinfo{volume}{VI}}},
  \bibinfo{pages}{197--206} (\bibinfo{year}{1867}).

\bibitem{Ranada1996}
\bibinfo{author}{Ra\~{n}ada, A.~F.} \& \bibinfo{author}{Trueba, J.~L.}
\newblock \bibinfo{title}{Ball lightning an electromagnetic knot?}
\newblock \emph{\bibinfo{journal}{Nature}}
  \href{http://dx.doi.org/10.1038/383032a0}{\textbf{\bibinfo{volume}{383}}},
  \bibinfo{pages}{32} (\bibinfo{year}{1996}).

\bibitem{Cruz2007}
\bibinfo{author}{Cruz, M.}, \bibinfo{author}{Turok, N.},
  \bibinfo{author}{Vielva, P.}, \bibinfo{author}{Mart\'{\i}nez-Gonz\'{a}lez,
  E.} \& \bibinfo{author}{Hobson, M.}
\newblock \bibinfo{title}{A cosmic microwave background feature consistent with
  a cosmic texture}.
\newblock \emph{\bibinfo{journal}{Science}}
  \href{http://dx.doi.org/10.1126/science.1148694}{\textbf{\bibinfo{volume}{318}}},
  \bibinfo{pages}{1612--1614} (\bibinfo{year}{2007}).

\bibitem{Smalyukh2009}
\bibinfo{author}{Smalyukh, I.~I.}, \bibinfo{author}{Lansac, Y.},
  \bibinfo{author}{Clark, N.~A.} \& \bibinfo{author}{Trivedi, R.~P.}
\newblock \bibinfo{title}{Three-dimensional structure and multistable optical
  switching of triple-twisted particle-like excitations in anisotropic fluids}.
\newblock \emph{\bibinfo{journal}{Nature Mater.}}
  \href{http://dx.doi.org/10.1038/nmat2592}{\textbf{\bibinfo{volume}{9}}},
  \bibinfo{pages}{139--145} (\bibinfo{year}{2009}).

\bibitem{Tkalec2011}
\bibinfo{author}{Tkalec, U.}, \bibinfo{author}{Ravnik, M.},
  \bibinfo{author}{Copar, S.}, \bibinfo{author}{Zumer, S.} \&
  \bibinfo{author}{Musevic, I.}
\newblock \bibinfo{title}{Reconfigurable knots and links in chiral nematic
  colloids}.
\newblock \emph{\bibinfo{journal}{Science}}
  \href{http://dx.doi.org/10.1126/science.1205705}{\textbf{\bibinfo{volume}{333}}},
  \bibinfo{pages}{62--65} (\bibinfo{year}{2011}).

\bibitem{Sec2014}
\bibinfo{author}{Se\v{c}, D.}, \bibinfo{author}{\v{C}opar, S.} \&
  \bibinfo{author}{\v{Z}umer, S.}
\newblock \bibinfo{title}{Topological zoo of free-standing knots in confined
  chiral nematic fluids}.
\newblock \emph{\bibinfo{journal}{Nat. Commun.}}
  \href{http://dx.doi.org/10.1038/ncomms4057}{\textbf{\bibinfo{volume}{5}}},
  \bibinfo{pages}{1--7} (\bibinfo{year}{2014}).

\bibitem{Han2010}
\bibinfo{author}{Han, D.}, \bibinfo{author}{Pal, S.}, \bibinfo{author}{Liu, Y.}
  \& \bibinfo{author}{Yan, H.}
\newblock \bibinfo{title}{Folding and cutting {DNA} into reconfigurable
  topological nanostructures}.
\newblock \emph{\bibinfo{journal}{Nature Nanotech.}}
  \href{http://dx.doi.org/10.1038/nnano.2010.193}{\textbf{\bibinfo{volume}{5}}},
  \bibinfo{pages}{712--717} (\bibinfo{year}{2010}).

\bibitem{Leach2005}
\bibinfo{author}{Leach, J.}, \bibinfo{author}{Dennis, M.~R.},
  \bibinfo{author}{Courtial, J.} \& \bibinfo{author}{Padgett, M.~J.}
\newblock \bibinfo{title}{Vortex knots in light}.
\newblock \emph{\bibinfo{journal}{New J. Phys.}}
  \href{http://dx.doi.org/10.1088/1367-2630/7/1/055}{\textbf{\bibinfo{volume}{7}}},
  \bibinfo{pages}{55} (\bibinfo{year}{2005}).

\bibitem{Dennis2010}
\bibinfo{author}{Dennis, M.~R.}, \bibinfo{author}{King, R.~P.},
  \bibinfo{author}{Jack, B.}, \bibinfo{author}{O'{H}olleran, K.} \&
  \bibinfo{author}{Padgett, M.~J.}
\newblock \bibinfo{title}{Isolated optical vortex knots}.
\newblock \emph{\bibinfo{journal}{Nature Phys.}}
  \href{http://dx.doi.org/10.1038/nphys1504}{\textbf{\bibinfo{volume}{6}}},
  \bibinfo{pages}{118--121} (\bibinfo{year}{2010}).

\bibitem{Kleckner2013}
\bibinfo{author}{Kleckner, D.} \& \bibinfo{author}{Irvine, W. T.~M.}
\newblock \bibinfo{title}{Creation and dynamics of knotted vortices}.
\newblock \emph{\bibinfo{journal}{Nature Phys.}}
  \href{http://dx.doi.org/10.1038/nphys2560}{\textbf{\bibinfo{volume}{9}}},
  \bibinfo{pages}{253--258} (\bibinfo{year}{2013}).

\bibitem{Kawaguchi2008}
\bibinfo{author}{Kawaguchi, Y.}, \bibinfo{author}{Nitta, M.} \&
  \bibinfo{author}{Ueda, M.}
\newblock \bibinfo{title}{Knots in a spinor {B}ose-{E}instein condensate}.
\newblock \emph{\bibinfo{journal}{Phys. Rev. Lett.}}
  \href{http://dx.doi.org/10.1103/PhysRevLett.100.180403}{\textbf{\bibinfo{volume}{100}}},
  \bibinfo{pages}{180403} (\bibinfo{year}{2008}).

\bibitem{Faddeev1997}
\bibinfo{author}{Faddeev, L.} \& \bibinfo{author}{Niemi, A.~J.}
\newblock \bibinfo{title}{Stable knot-like structures in classical field
  theory}.
\newblock \emph{\bibinfo{journal}{Nature}}
  \href{http://dx.doi.org/10.1038/387058a0}{\textbf{\bibinfo{volume}{387}}},
  \bibinfo{pages}{58--61} (\bibinfo{year}{1997}).

\bibitem{Battye1998}
\bibinfo{author}{Battye, R.~A.} \& \bibinfo{author}{Sutcliffe, P.~M.}
\newblock \bibinfo{title}{Knots as stable soliton solutions in a
  three-dimensional classical field theory}.
\newblock \emph{\bibinfo{journal}{Phys. Rev. Lett.}}
  \href{http://dx.doi.org/10.1103/PhysRevLett.81.4798}{\textbf{\bibinfo{volume}{81}}},
  \bibinfo{pages}{4798--4801} (\bibinfo{year}{1998}).

\bibitem{Nakahara2003}
\bibinfo{author}{Nakahara, M.}
\newblock \emph{\bibinfo{title}{Geometry, Topology and Physics}}
  (\bibinfo{publisher}{Taylor \& Francis Group}, \bibinfo{address}{Boca Raton},
  \bibinfo{year}{2003}).

\bibitem{Hopf1931}
\bibinfo{author}{Hopf, H.}
\newblock \bibinfo{title}{\"{U}ber die {A}bbildungen der dreidimensionalen
  {S}ph\"are auf die {K}ugelfl\"ache}.
\newblock \emph{\bibinfo{journal}{Mathematische Annalen}}
  \href{http://dx.doi.org/10.1007/bf01457962}{\textbf{\bibinfo{volume}{104}}},
  \bibinfo{pages}{637--665} (\bibinfo{year}{1931}).

\bibitem{Urbantke2003}
\bibinfo{author}{Urbantke, H.}
\newblock \bibinfo{title}{The {H}opf fibration---seven times in physics}.
\newblock \emph{\bibinfo{journal}{J. Geom. Phys.}}
  \href{http://dx.doi.org/10.1016/s0393-0440(02)00121-3}{\textbf{\bibinfo{volume}{46}}},
  \bibinfo{pages}{125--150} (\bibinfo{year}{2003}).

\bibitem{Moore2008}
\bibinfo{author}{Moore, J.~E.}, \bibinfo{author}{Ran, Y.} \&
  \bibinfo{author}{Wen, X.-G.}
\newblock \bibinfo{title}{Topological surface states in three-dimensional
  magnetic insulators}.
\newblock \emph{\bibinfo{journal}{Phys. Rev. Lett.}}
  \href{http://dx.doi.org/10.1103/PhysRevLett.101.186805}{\textbf{\bibinfo{volume}{101}}},
  \bibinfo{pages}{186805} (\bibinfo{year}{2008}).

\bibitem{Hietarinta2004}
\bibinfo{author}{Hietarinta, J.}, \bibinfo{author}{J\"aykk\"a, J.} \&
  \bibinfo{author}{Salo, P.}
\newblock \bibinfo{title}{Relaxation of twisted vortices in the
  {F}addeev-{S}kyrme model}.
\newblock \emph{\bibinfo{journal}{Phys. Lett. A}}
  \href{http://dx.doi.org/10.1016/j.physleta.2003.11.042}{\textbf{\bibinfo{volume}{321}}},
  \bibinfo{pages}{324--329} (\bibinfo{year}{2004}).

\bibitem{Barenghi2007}
\bibinfo{author}{Barenghi, C.~F.}
\newblock \bibinfo{title}{Knots and unknots in superfluid turbulence}.
\newblock \emph{\bibinfo{journal}{Milan. J. Math.}}
  \href{http://dx.doi.org/10.1007/s00032-007-0069-5}{\textbf{\bibinfo{volume}{75}}},
  \bibinfo{pages}{177--196} (\bibinfo{year}{2007}).

\bibitem{Manton2004}
\bibinfo{author}{Manton, N.} \& \bibinfo{author}{Sutcliffe, P.}
\newblock \emph{\bibinfo{title}{Topological Solitons}}
  (\bibinfo{publisher}{Cambridge University Press}, \bibinfo{address}{New
  York}, \bibinfo{year}{2004}).

\bibitem{Radu2008}
\bibinfo{author}{Radu, E.} \& \bibinfo{author}{Volkov, M.~S.}
\newblock \bibinfo{title}{Stationary ring solitons in field theory --- knots
  and vortons}.
\newblock \emph{\bibinfo{journal}{Phys. Rep.}}
  \href{http://dx.doi.org/10.1016/j.physrep.2008.07.002}{\textbf{\bibinfo{volume}{468}}},
  \bibinfo{pages}{101--151} (\bibinfo{year}{2008}).

\bibitem{Ranada1990}
\bibinfo{author}{Ra\~{n}ada, A.~F.}
\newblock \bibinfo{title}{Knotted solutions of the {M}axwell equations in
  vacuum}.
\newblock \emph{\bibinfo{journal}{J. Phys. A: Math. Gen.}}
  \href{http://dx.doi.org/10.1088/0305-4470/23/16/007}{\textbf{\bibinfo{volume}{23}}},
  \bibinfo{pages}{L815--L820} (\bibinfo{year}{1990}).

\bibitem{Denschlag2000}
\bibinfo{author}{Denschlag, J.} \emph{et~al.}
\newblock \bibinfo{title}{Generating solitons by phase engineering of a
  {B}ose-{E}instein condensate}.
\newblock \emph{\bibinfo{journal}{Science}}
  \href{http://dx.doi.org/10.1126/science.287.5450.97}{\textbf{\bibinfo{volume}{287}}},
  \bibinfo{pages}{97--101} (\bibinfo{year}{2000}).

\bibitem{Burger1999}
\bibinfo{author}{Burger, S.} \emph{et~al.}
\newblock \bibinfo{title}{Dark solitons in {B}ose-{E}instein condensates}.
\newblock \emph{\bibinfo{journal}{Phys. Rev. Lett.}}
  \href{http://dx.doi.org/10.1103/PhysRevLett.83.5198}{\textbf{\bibinfo{volume}{83}}},
  \bibinfo{pages}{5198--5201} (\bibinfo{year}{1999}).

\bibitem{Vinen1961}
\bibinfo{author}{Vinen, W.~F.}
\newblock \bibinfo{title}{The detection of single quanta of circulation in
  liquid helium {II}}.
\newblock \emph{\bibinfo{journal}{Proc. R. Soc. A}}
  \textbf{\bibinfo{volume}{260}}, \bibinfo{pages}{218--236}
  (\bibinfo{year}{1961}).

\bibitem{Choi2012}
\bibinfo{author}{Choi, J.-y.} \emph{et~al.}
\newblock \bibinfo{title}{Imprinting skyrmion spin textures in spinor
  {B}ose-{E}instein condensates}.
\newblock \emph{\bibinfo{journal}{New J. Phys.}} \textbf{\bibinfo{volume}{14}},
  \bibinfo{pages}{053013} (\bibinfo{year}{2012}).

\bibitem{Ray2015}
\bibinfo{author}{Ray, M.~W.}, \bibinfo{author}{Ruokokoski, E.},
  \bibinfo{author}{Tiurev, K.}, \bibinfo{author}{M\"ott\"onen, M.} \&
  \bibinfo{author}{Hall, D.~S.}
\newblock \bibinfo{title}{Observation of isolated monopoles in a quantum
  field}.
\newblock \emph{\bibinfo{journal}{Science}}
  \href{http://dx.doi.org/10.1126/science.1258289}{\textbf{\bibinfo{volume}{348}}},
  \bibinfo{pages}{544--547} (\bibinfo{year}{2015}).

\bibitem{Ray2014}
\bibinfo{author}{Ray, M.~W.}, \bibinfo{author}{Ruokokoski, E.},
  \bibinfo{author}{Kandel, S.}, \bibinfo{author}{M\"ott\"onen, M.} \&
  \bibinfo{author}{Hall, D.~S.}
\newblock \bibinfo{title}{Observation of {D}irac monopoles in a synthetic
  magnetic field}.
\newblock \emph{\bibinfo{journal}{Nature}}
  \href{http://dx.doi.org/10.1038/nature12954}{\textbf{\bibinfo{volume}{505}}},
  \bibinfo{pages}{657--660} (\bibinfo{year}{2014}).

\end{thebibliography}



\textbf{Acknowledgements} We acknowledge funding by the National Science Foundation (grant PHY--1205822), by the Academy of Finland through its Centres of Excellence Program (grant no.\ 251748) and grants (nos\ 135794 and 272806), Finnish Doctoral Programme in Computational Sciences, and the Magnus Ehrnrooth Foundation. CSC - IT Center for Science Ltd.\ (Project No.\ ay2090) and Aalto Science-IT project are acknowledged for computational resources. We thank N. Johnson for making public his Hopf fibration code, A. Li for assistance with figures, and W. Lee and S.J. Vickery for experimental assistance.

\textbf{Author Contributions} M.W.R., A.H.G, and D.S.H.\ developed and conducted the experiments and analysed the data. K.T. and E.R. performed the numerical simulations under the guidance of M.M. who provided the initial suggestions for the experiment. M.M. and D.S.H. developed the analytical interpretation of the $m=0$ data as preimages. All authors discussed both experimental and theoretical results and commented on the manuscript.

\textbf{Author Information} 
The authors declare that they have no competing financial interests. Correspondence and requests for materials should be addressed to D.S.H.\ (dshall@amherst.edu).

\textbf{*PLEASE NOTE} This is the version originally submitted to the journal and contains some minor typos and errors. The published and corrected version will be posted to arXiv six months after publication in the journal.






\clearpage

\setcounter{page}{1}
\renewcommand*{\thepage}{ED--\arabic{page}}

\stepcounter{myfigure}
\newcommand{\myfont}[1]{}
\renewcommand{\figurename}{Extended Data Figure}

\end{document}